\documentclass[a4paper,11pt]{article}

\usepackage{amsmath}
\usepackage{amsfonts}
\usepackage{amssymb}
\usepackage{bbm}
\usepackage{pstricks}
\usepackage{pst-plot}
\usepackage{pst-node}
\usepackage{multido}
\usepackage[all]{xy}
\usepackage{rotating}
\usepackage[dvips]{epsfig}
\usepackage{colortbl}
\usepackage{wrapfig}
\usepackage[hang,nooneline]{subfigure}
\usepackage[footnotesize,ruled]{caption}

\usepackage[utf8]{inputenc}

\usepackage{euscript}

\begin{document}

\title{Confronting Finsler space--time with experiment}
\author{Claus L\"ammerzahl, Dennis Lorek, Hansj\"org Dittus$^*$ \\
ZARM, University of Bremen, Am Fallturm, 28359 Bremen, Germany \\ 
$^*$ now at German Aerospace Center, Institute of Space Systems, Bremen}

\maketitle

\begin{abstract}
Within all approaches to quantum gravity small violations of the Einstein Equivalence Principle are expected. This includes violations of Lorentz invariance. While usually violations of Lorentz invariance are introduced through the coupling to additional tensor fields, here a Finslerian approach is employed where violations of Lorentz invariance are incorporated as an integral part of the space--time metrics. Within such a Finslerian framework a modified dispersion relation is derived which is confronted with current high precision experiments. As a result, Finsler type deviations from the Minkowskian metric are excluded with an accuracy of $10^{-16}$.
\end{abstract}

\section{Introduction}

The Einstein Equivalence Principle implies that gravity has to be a metric theory \cite{Will93,Laemmerzahl06}. This means that gravity has to be mathematically described by a pseudo--Riemannian space--time metric. For all metric theories one can calculate all the standard effects like light deflection, perihelion shift, gravitational time delay, etc. These effects take certain values if the metric is related to the masses in the universe according to Einsteins field equation. All gravitational phenomena (except perhaps dark matter and dark energy) are fully describable within standard Einsteins General Relativity (GR).

It is, however, generally accepted that due to the incompatibility between GR and Quantum Theory there should be a more complete theory called Quantum Gravity. In the low energy and/or classical limit this theory should reproduce GR with small corrections. These small corrections necessarily violate the Einstein Equivalence Principle \cite{Laemmerzahl06}.

There are many possibilities to incorporate modifications of GR. One way is to include violations of the Universality of Free Fall and the Universality of the Gravitational Redshift. This usually comes along with scalar--tensor theories emerging from the low energy limit of string theory. Another way is to break Lorentz invariance through a coupling with tensor fields as they appear from spontaneous symmetry breaking of string scenarios. This is done phenomenologically in the Standard Model Extension \cite{ColladayKostelecky97} or in more general frameworks \cite{LaemmerzahlMaciasMueller05}. A further way is to include other dynamical tensor fields like torsion \cite{Hehletal76}. On the kinematical level violations of Lorentz invariance have been treated within the Robertson--Mansouri--Sexl framework \cite{Robertson49,MansouriSexl77}.

Here we are proposing another scheme of breaking Lorentz invariance by generalizing the space--time metric to be of Finslerian form. While in all scenarios mentioned above Lorentz invariance is broken by introducing additional fields while leaving the metric in a Lorentzian form, within a Finslerian framework it is the metric itself which breaks Lorentz invariance. In this framework it is not possible to find a coordinate system so that the metric acquires a Minkowskian form. A Finslerian anisotropy does not vanish by considering infinitesimal distances. By allowing the Finsler metric to depend on space and time, a Finsler geometry can been taken as a wider framework to study the structure of space--time and gravitation, and to analyze space experiments \cite{Roxburgh92}. Such a modification of GR has found some interest among theorists.

Finsler space--times have been discussed in \cite{GibbonsGomisPope2007} showing that a deformation of very special relativity, that produces a curved space--time with a cosmological constant, leads in a natural way to a Finsler geometry. Moreover, in \cite{GirelliLiberatiSindoni2007} it is shown that the common feature of all Quantum Gravity phenomenology approaches, considering a modification of the mass--shell condition for relativistic particles, corresponds in a certain way to a Finsler geometry. In the 1950s, Finsler metrics were discussed in the search for a unified theory of electromagnetic and gravitational fields \cite{KilmisterStephenson54}. More recently, the possibility of a gravitational theory within a Finslerian framework has been explored, that would possibly lead to modifications to the observational results as predicted by GR, see \cite{Roxburgh92,Rutz93} and references cited therein. An extensive list of pre--1985 literature can be found in Asanov's book on the subject \cite{Asanov85}.

As already mentioned, Finslerian types of space--time metrics usually are not included in schemes aiming at describing or parametrizing violations of Lorentz invariance. Here we like to set up a Finslerian frame of violations of Lorentz invariance and confront it with current high precision experiments. We base our approach to that class of Finslerian metrics which is obtained from the direct measurement of the speed of light which is allowed to depend on the direction. A recent review of indefinite Finsler spaces appropriate for relativistic problems has been given by Perlick \cite{Perlick05}. The class of Finsler space--times discussed in the present paper fits into the class given in his Example 3.3. In this framework we derive the modified dispersion relation for the propagation of electromagnetic waves and compare its predictions with null--results of Michelson--Morley experiments. Such a comparison is lacking so far in the literature and gives new precise constraints on the Lorentz invariance violating parameters.

A further motivation to consider Finsler space--times is the following: In GR, the distance between two space--time events as well as the gravitational field governing the motion of particles are given by a quadratic form, the Riemannian metric $g_{\mu\nu}$. This metric can be transformed pointwise to the Minkowskian form $\eta_{\mu\nu} = \hbox{diag}(+\, -\, -\, -)$. This can be justified within a constructive axiomatic approach to the Riemannian structure of space--time \cite{EhlersPiraniSchild72}. This approach, however, requires the differentiability of a two--point function representing the geodetic distance. Releasing this requirements would allow Finslerian structures instead of Riemannian structures as derived in \cite{EhlersPiraniSchild72}.

\section{Finsler Geometry}

Riemann \cite{Riemann1854} himself questioned whether distances in space have to be described by a quadratic form and discussed, instead, a quartic form given by $ds^4 = g_{\mu\nu\rho\sigma}(x) dx^\mu  dx^\nu  dx^\rho  dx^\sigma$. A wider framework including this example is given by Finsler spaces \cite{Finsler18,Rund53} in which the length interval $ds$ is given by a general metric function $F(x,\,dx)$ of the coordinate increments $dx^\mu$ which is homogeneous of degree one in the $dx^\mu$
\begin{equation}
ds^2 = F^2(x,\,dx) \, , \qquad F(x,\, \lambda dx) = \lambda F(x,\,dx) \, . \label{GenFinsler}
\end{equation}
The Finsler metric tensor $f_{\mu\nu}(x,\,dx)$ is defined as
\begin{equation}
ds^2 = f_{\mu\nu}(x,\,dx) dx^\mu dx^\nu \, , \quad \text{where} \quad  f_{\mu\nu}(x,\,y) = \frac{1}{2} \frac{\partial^2 F^2(x^k,\,y^m)}{\partial y^\mu \partial y^\nu} \,.
\end{equation}
This formalism, however, in general is well defined for positive definite metrics only. In this case the geodesic equation can as usual be derived from the extremal $\delta \int ds = 0$. The special case of a quartic metric $ds^4 = g_{\mu\nu\rho\sigma}(x) dx^\mu dx^\nu dx^\rho dx^\sigma$ already considered by Riemann is only one example of the class of ``power law'' metrics
\begin{equation}
ds = (g_{\mu_1 \mu_2 \ldots \mu_n}(x) dx^{\mu_1} dx^{\mu_2} \cdots dx^{\mu_n})^{1/n} \, , \label{FinslerPowerLaw}
\end{equation}
where $g_{\mu_1 \mu_2 \ldots \mu_n}(x)$ is a covariant tensor field of rank $n$. 

For a particular indefinite case \cite{Roxburgh92} this wider framework was considered to examine the post--Newtonian limit for the spherically symmetric one--body problem assuming that the metric reduces in the absence of a gravitational field to the Minkowski space of Special Relativity (SR). It has been shown \cite{Roxburgh92,Coley82,Roxburgh91} that such a metric is compatible with standard solar system tests. In contrast, here we will consider small Quantum Gravity motivated perturbations in the metric, which remain even in the absence of a gravitational field.

According to the definite case, also in the general indefinite case the motion of particles and light rays are  incorporated into a Finslerian framework for the gravitational interaction by postulating -- as in Riemannian geometry -- that particles and light rays are extremals of the functional $\int ds$ (where for light we have in addition to require $ds=0$). For a general indefinite Finsler metric of type (\ref{GenFinsler}) or (\ref{FinslerPowerLaw}) this is, however, not admissible \cite{Beem70}. Instead, one has to restrict to a certain subclass of Finslerian metrics \cite{Perlick05,Beem70} given by $ds^2 = F(x, dx)$, where $F$ is homogeneous of degree two, $F(x, \lambda dx) = \lambda^2 F(x, dx)$ so that $ds^2 = g_{\mu\nu}(x, dx) dx^\mu dx^\nu$ with $g_{\mu\nu}(x, y) = \frac{\partial^2 F(x, y)}{\partial y^\mu \partial y^\nu}$. The physically admissible Finslerian metrics we are going to introduce below through the measurement of the velocity of light is of this class. 

\section{Finsler Metric and Light Propagation}

In SR the velocity of light is given by
\begin{equation}
c^2 = \frac{\delta_{ij} dx^i dx^j}{(dt)^2} \label{VelLight}
\end{equation}
($i, j = 1,\,2,\,3$), which is equivalent to $c^2 (dt)^2 = \delta_{ij} dx^i dx^j$ and which leads to
\begin{equation}
0 = c^2 dt^2 - \delta_{ij} dx^i dx^j = \eta_{ab} dx^a dx^b \, , \label{LocalLorentzMetric}
\end{equation}
where $a, b = 0, \ldots, 3$ and $\eta_{ab}$ is the Minkowski metric. The main point here is that for a given $dt$ the set of points fulfilling $c^2 dt^2 = \delta_{ij} dx^i dx^j$ gives a sphere of radius $dr = \sqrt{\delta_{ij} dx^i dx^j}$. Even if the condition $ds^2 = c^2 dt^2 - \delta_{ij} dx^i dx^j = 0$ is replaced by a more general one, namely $ds^2 = g_{ab} dx^a dx^b=0$, one always can find a coordinate system so that \eqref{LocalLorentzMetric} holds.

When we now consider small Quantum Gravity motivated perturbations in the metric, even in the absence of a gravitational field, \eqref{VelLight} has to be replaced by
\begin{equation}
c^2 = \frac{D(dx^i)}{(dt)^2}  \, ,  \label{VelLightF}
\end{equation}
where $D(dx)$ is a function that is homogeneous of degree 2 and not reducible to a quadratic form. In that case no coordinate system can be found so that \eqref{LocalLorentzMetric} holds. Condition \eqref{VelLightF} then can be reformulated as
\begin{equation}
0 = c^2 dt^2 - D(dx^i) =: G(dx^i) \, ,
\end{equation}
where $G(dx^i)$ constitutes a Finslerian metric $ds^2 = G(dx^i)$. A comparison with Example 3.3 in \cite{Perlick05} confirms that this form belongs to the physically most relevant class of indefinite Finsler space--times.

In the following we like to confront this type of Finslerian metrics with Michelson--Morley type experiments. In order to do so we have to introduce a kind of parametrization of the deviation of Finslerian metrics from the standard Minkowski metrics. This can be done most conveniently provided $D(dx^i)$ is based on a tensor of rank $2p$: $D^p(dx^i) = D_{i_1 \cdots i_{2p}} dx^{i_1} \cdots dx^{i_{2p}}$. In these cases a deviation from the Minkowski case can be easily parametrized as
\begin{equation}
D^p(dx^i) = D_{i_1 \cdots i_{2p}} dx^{i_1} \cdots dx^{i_{2p}} = \left(\delta_{ij} dx^i dx^j\right)^p + \phi_{i_1 \cdots i_{2p}} dx^{i_1} \cdots dx^{i_{2p}} \, .
\end{equation}
Any deviation from a Minkowski space--time is encoded in the coefficients $\phi_{i_1 \cdots i_{2p}}$. For $\phi_{i_1 \cdots i_{2p}} = 0$ we recover Minkowski space--time. Since until now no deviation from Lorentz symmetry has been found we assume in the following that $\phi_{i_1 \cdots i_{2p}} \ll 1$. Therefore, our Finslerian metric reads
\begin{equation}\label{start}
ds^2 = (c\,dt)^2 - D(dx^i) = (c\,dt)^2 - \delta_{ij} dx^i dx^j \left(1 + \frac{1}{p} \frac{\phi_{i_1 \cdots i_{2p}} dx^{i_1} \cdots dx^{i_{2p}}}{\left(\delta_{ij} dx^i dx^j\right)^p}\right) + {\cal O}(\phi^2) = G(dx^i) \, .
\end{equation}
This type of Finslerian metric fits into the class given in \cite{Perlick05}. Moreover, the spatial part of our Finslerian metric is of Berwald--Moor type \cite{BaoChernShen2000} as it is given by a polynomial of degree $2p$.

\section{Modified Dispersion Relation}\label{section:ModDispRel}

Based on the above Finslerian metric we derive now the relation between the frequency $\omega$ and the wave vector $k_i$ of a light ray $l$ fulfilling $G(l) = 0$, that is, we derive the corresponding dispersion relation. This is needed because the description of the experiments are based on frequencies and wave vectors and, thus, on the dispersion relation. Since this is a purely local procedure we omit the reference to a particular space--time point $x$.

We start with (\ref{start}) written as
\begin{equation}\label{Fgeneral}
G(l) = (l^0)^2 - l^2 \left(1 + \frac{1}{p} \frac{\phi_{i_1 \hdots i_{2p}} l^{i_1} \hdots l^{i_{2p}}}{l^{2p}} \right) = 0 \, ,
\end{equation}
where $l^2 = \delta_{ij}l^i l^j$. As
\begin{equation}
{\omega/c \choose k_i} = \frac{1}{2} {\partial G / \partial l^0 \choose \partial G / \partial l^i }\,,
\end{equation}
the frequency $\omega$ is given by $\omega = cl^0$ and the wave vector $k_i$ can be written as
\begin{equation}
k_i = - \delta_{ij}l^j + \frac{p-1}{p} \frac{\delta_{ij}l^j  \phi_{i_1 \hdots i_{2p}} l^{i_1} \hdots l^{i_{2p}}}{l^{2p}} - \frac{\phi_{i_1 \hdots i_{2p}} \delta^{i_1}_i l^{i_2} \hdots l^{i_{2p}} }{l^{2p-2}} \, ,
\end{equation}
that has to be solved for $l=l(k)$ with help of the Ansatz $l^i = l^i_{(0)} + l^i_{(1)} \hdots$, where $l^i_{(0)}$ is given by $l^i_{(0)} = - \delta^{ij} k_j$ and $l^i_{(1)}$ is of the order of $\phi_{i_1 \ldots i_{2p}}$ and spatial indices are raised with $\delta^{ij}$.
This gives
\begin{equation}\label{lgeneral}
l^i = - \delta^{ij} k_j \left(1 + \frac{p-1}{p} \frac{\phi^{i_1 \hdots i_{2p}} k_{i_1} \hdots k_{i_{2p}} }{k^{2p}}\right) + \frac{\phi^{i_1 \hdots i_{2p}} \delta_{i_1}^i k_{i_2} \hdots k_{i_{2p}} }{k^{2p-2}} \,,
\end{equation}
with $k^2 = \delta^{ij} k_i k_j$. When we insert $\omega = cl^0$ and (\ref{lgeneral}) into (\ref{Fgeneral}), we obtain
\begin{equation}
\omega^2 = c^2 k^2 \left(1 - \frac{1}{p} \frac{\phi^{i_1 \hdots i_{2p}} k_{i_1} \hdots k_{i_{2p}}}{k^{2p}}\right)
\end{equation}
or
\begin{equation}
\omega^2 = c^2k^2 \left( 1 - \frac{1}{p}  \phi^{i_1 \hdots i_{2p}} n_{i_1} \hdots n_{i_{2p}} \right) \, ,
\end{equation}
where the components of the direction of wave propagation are given by $n_i := k_i/|k|$. For a propagation in the $x$--$y$--plane, i.e. $n_z=0$, we obtain the general modified dispersion relation
\begin{equation}\label{disp1general}
\omega^2 = c^2 k^2 \left(1 - \frac{1}{p} {\cal K}(\varphi)\right) \,,
\end{equation}
with
\begin{equation}
{\cal K}(\varphi) =  \sum_{n=0}^{2p} \frac{(2p)!}{n!(2p-n)!} \phi^{[n]} \cos^{2p -n} \varphi \, \sin^n\varphi \,,
\end{equation}
where $\varphi$ defines the angle between the $x$--axis and the direction of propagation, and the index $n$ in the tensor $\phi^{[n]}$ denotes the number of indices that correspond to the $y = x^2$--direction, that is, $\phi^{[0]} = \phi^{111\hdots1}$, $\phi^{[1]} = \phi^{211\hdots1}$, $\phi^{[2]} = \phi^{221\hdots1}$ and so on. The dispersion relation \eqref{disp1general} relates the frequency $\omega$ to a given wave vector $k_i$. With an appropriate coordinate transformation we may choose $\phi^{[0]} = 0$. Then
\begin{equation}
{\cal K}(\varphi) =  \sum_{n=1}^{2p} \frac{(2p)!}{n!(2p-n)!} \phi^{[n]} \cos^{2p -n} \varphi \, \sin^n\varphi \,. \label{DefK}
\end{equation}

In the case $p=1$ we obtain
\begin{equation}\label{dispnew1}
\omega^2 = c^2 k^2 \left[1 - \left(2 \phi^{12} \cos\varphi \, \sin\varphi + \phi^{22} \sin^2\varphi\right)\right] \, .
\end{equation}
By means of a coordinate transformation the quantities $\phi^{ij}$ can be made to  vanish. For a quartic metric, that is $p = 2$, the modified dispersion relation (\ref{disp1general}) yields
\begin{eqnarray}\label{disp1}
\omega^2 & = & c^2k^2 \Biggl[1- \frac{1}{2} \Bigl(4 \phi^{1112} \cos^3\varphi \, \sin\varphi + 6 \phi^{1122} \cos^2\varphi \, \sin^2\varphi \nonumber\\
& & \phantom{c^2k^2 \Biggl[1- \frac{1}{2} \Bigl(}  + 4 \phi^{1222} \cos\varphi \, \sin^3\varphi + \phi^{2222} \sin^4\varphi \Bigr) \Biggr] \,.
\end{eqnarray}
Here, the $\phi^{i_1 \ldots i_4}$ cannot be made to vanish by means of a coordinate transformation.

\section{Comparison with Experiments}

We now would like to confront the derived modified dispersion relation (\ref{disp1general}) with current high precision experiments testing Lorentz invariance. These experiments use optical resonators with a length $L$.

In one approach this length $L$ is just the length operationally {\it defined} by the resonator itself. This length can in principle behave like a Finslerian length based on Finslerian metric different from the Finslerian metric governing the propagation of light. As a consequence, by means of Michelson--Morley experiments we can explore only whether these two Finslerian metrics coincide or not.

In another approach we may consider the optical resonator as being made of atoms which obey a quantum equation which is effectively based on a possibly generalized Dirac equation \cite{AudretschLaemmerzahl93}. Therefore the length $L$ and, thus, the metric defined by the optical resonator is an outcome of the properties of the Dirac equation. Within a kind of a constructive axiomatic approach using elements of quantum mechanics \cite{AudretschLaemmerzahl93} it has been shown that quantum particles with spin which define only two light cones and possess only two spin states necessarily define a Riemannian metric only. If one takes these assumptions as experimentally granted, then the metric defined by the length of the resonator can also be a Riemannian one only\footnote{In principle, one also should calculate the modification of the length of the cavity induced by the modified dispersion relation. This, however, is not necessary for standard situations: it has been shown in \cite{Muelleretal2003} that such kind of modification can be neglected for materials currently used in such type of cavity experiments.}. Therefore, in this case the presently discussed Michelson--Morley experiments show whether light propagates according to a Riemannian or a Finslerian metric.

Due to the boundary conditions there are standing waves inside the optical resonators with a frequency given by the dispersion relation where the modulus of the wave vector is given by $k = \frac{m}{L}$, where $m$ is the mode number. Turning around the cavity will result in a change of the frequency according to the orientation dependence of the function ${\cal K}(\varphi)$ appearing in the dispersion relation$^1$.

The most precise Michelson--Morley experiment searching for an anisotropy of the speed of light has been carried through by M\"uller \textit{et al.} \cite{Muelleretal2007}. They correlated the data of measurements of two optical resonators (in Berlin, Germany, and in Perth, Australia) over a period of more than $1$\,yr. Their constraints for the various Fourier components are presented in Fig.\,2 in \cite{Muelleretal2007}. These null--results exclude a $\varphi$--dependence of the modified dispersion relation with an accuracy of $10^{-16}$. Therefore, the quantity ${\cal K}(\varphi)$ has to be smaller than $10^{-16}$.

Assuming that ${\cal K}(\varphi) \equiv 0$ exactly, also all derivatives of Eqn.\,(\ref{disp1general}) with respect to the angle $\varphi$ have to vanish. The first derivative of $\mathcal{K}$ with respect to $\varphi$ yields
\begin{equation}
\frac{d\mathcal{K}}{d\varphi} = \sum \limits_{n=1}^{2p} \binom{2p}{n}   \phi^{[n]}\left(n \cos^{2p -n+1}\varphi \, \sin^{n-1}\varphi - (2p-n)  \cos^{2p -n-1}\varphi \, \sin^{n+1}\varphi\right) \, . \label{Kfirstderivative}
\end{equation}
This should vanish for all $\varphi$. If we choose $\varphi=0$, then this equation obviously implies $\phi^{[1]}=0$. Therefore, in \eqref{DefK} and \eqref{Kfirstderivative} the sum effectively starts with $n = 2$. Differentiating ${\cal K}$ twice and setting again $\varphi = 0$ yields the condition $\phi^{[2]} = 0$. By induction, we can proceed showing that $\phi^{[n]} = 0$ for all $0 < n \leq 2p$. Therefore all parameters $\phi_{i_1 \ldots i_{2p}}$ vanish and the space--time becomes Minkowskian. As a consequence Finsler metrics can be excluded with an accuracy of $10^{-16}$.

\section{Conclusion}

All Quantum Gravity approaches lead to small modifications of GR violating the Einstein Equivalence Principle. One possibility is a violation of Lorentz invariance. Schemes aiming at describing or parametrizing dynamical violations of Lorentz invariance include the Standard Model Extension and generalizations of it. All these test theories do not include Finslerian modifications of SR or GR. In this paper we set up a Finslerian frame of Lorentz invariance violation which we confronted with current high precision experiments testing Lorentz invariance. We introduced an appropriate parametrization of the deviation of Finslerian metrics from the standard Minkowski metrics. Based on that we derived a modified dispersion relation for the propagation of electromagnetic waves where the frequency depends on the direction of propagation. Utilizing the null--results of the currently most precise Michelson--Morley experiments enabled us to constrain the modifying parameters in the dispersion relation by factors with magnitudes of $10^{-16}$. By reason of these results Finsler metrics can be excluded with an accuracy of $10^{-16}$.

\section{Acknowledgments}

C.\,L. would like to thank the German Aerospace Center DLR for financial support. C.\,L. and D.\,L. would also like to thank the German Research Foundation and the Centre for Quantum Engineering and Space--Time Research QUEST for financial support.

\end{document}